# Toward an Integrated Framework for Automated Development and Optimization of Online Advertising Campaigns

STAMATINA THOMAIDOU, Athens University of Economics and Business
MICHALIS VAZIRGIANNIS, Athens University of Economics and Business &LIX, Ecole Polytechnique, France
KYRIAKOS LIAKOPOULOS, Athens University of Economics and Business

Creating and monitoring competitive and cost-effective pay-per-click advertisement campaigns through the web-search channel is a resource demanding task in terms of expertise and effort. Assisting or even automating the work of an advertising specialist will have an unrivaled commercial value. In this paper we propose a methodology, an architecture, and a fully functional framework for semi- and fully- automated creation, monitoring, and optimization of cost-efficient pay-per-click campaigns with budget constraints. The campaign creation module generates automatically keywords based on the content of the web page to be advertised extended with corresponding ad-texts. These keywords are used to create automatically the campaigns fully equipped with the appropriate values set. The campaigns are uploaded to the auctioneer platform and start running. The optimization module focuses on the *learning process* from existing campaign statistics and also from applied strategies of previous periods in order to invest optimally in the next period. The objective is to maximize the performance (i.e. clicks, actions) under the current budget constraint. The fully functional prototype is experimentally evaluated on real world Google AdWords campaigns and presents a promising behavior with regards to campaign performance statistics as it outperforms systematically the competing manually maintained campaigns.

Categories and Subject Descriptors: H.4.m [**Information Systems Applications**]: Miscellaneous

General Terms: Algorithms, Design, Experimentation, Measurement

Additional Key Words and Phrases: online advertising, pay-per-click advertising, automated campaign management, textual advertising, keyword selection, ad creative, genetic algorithms, Google AdWords

## 1. INTRODUCTION AND MOTIVATION

### 1.1. Online Advertising Campaigns

Online advertising is gaining acceptance and market share while it has evolved into a $26 billion industry for advertisers [1]. One form of online advertising is the promotion of products and services through search-based advertising. This follows an ad auction process with pay-per-click (PPC) model [Douzet ] for the advertisers. The selected search engine has the role of the auctioneer for the ad slots and the bids that the advertisers set for keywords and their ads. The dominant strategy for ad selection is the hybrid second-price auction [Edelman et al. 2007] system. The three most prevalent options in the search-based advertising market are Google AdWords, Yahoo Search Marketing, and Microsoft AdCenter (the two latter have merged) [2]. Today's most popular search-based advertising platform is Google AdWords having the largest share of revenues amongst its competitors. Google [3] 2010 annual report showed that company's advertising revenues made up 97% of its revenues in 2008 and 2009, and 96% of its revenues in 2010. In 2010, Google advertising revenues (as an auctioneer) were $ 29 billion [4]. Search remains the largest online advertising revenue format, accounting for 46.5% of 2011 revenues, up from 44.8% in 2010. In 2011, Search revenues

---

totaled $14.8 billion, up almost 27% from $11.7 billion in 2010. Search has remained the leading format since 2006, having strong sequential growth. [5].

Effective *keyword selection* is one of the most important success factors for online advertising. Companies would like to advertise on the most effective keywords to attract only prospective customers and not uninterested browsing users. In addition, they need well-written ad creatives to attract more visitors and generate thus higher revenues.

The major problem of an advertising campaign that takes into account only the suggestions of the most popular terms is that they are widely used, therefore the relevant keywords are quite competitive in terms of cost-per-click (CPC) cost. Another issue is that they are volume-based (i.e. number of monthly searches), which means these keywords will tend to drive more traffic to the campaign but not necessarily proportional conversions.

Thus, to tackle the above issues we propose extracting keywords from a given landing page and then extend them with further terms that are highly relevant yet not-obvious to the given input keyword with less competition in terms of CPC. Also, in order to facilitate the process of the ad text creation, our system generates automatically for each landing page an ad creative using text summarization.

In addition, the preparation of large scale online advertising campaigns for products, services, brands, or web pages can be a very complex task if it is designed for websites with online catalogs or catalog aggregators. The shops or listings are classified according to the products that they are selling, so each landing page contains important information and relevant description for each category or product that needs to be considered. The number of the various urls inside these domains makes the effort even more complicated regarding the manual insertion of keywords and ad-texts per landing page. Our proposed system aims at the automation of this procedure in order to aid the advertisers.

## 1.2. Contributions

This paper offers an integrated approach and a fully functioning prototype system for automated advertising campaign creation, management, and monitoring for profit optimization under budget constraints. Thus during the campaign management process, we focus also on budget optimization - a very challenging issue - presenting a methodology for selecting the most effective keywords and their bids, aiming at maximizing either the profits for the advertiser based on a specific budget or the traffic on their website. In this effort, we focus only on the advertisers and not on the other bidders or the self-interested auctioneer. We also assume that there are no public data on the competitors bidding behavior. Nevertheless, the auctioneer, which in our case is Google AdWords, provides us with two important variables: *Global Monthly Searches* and *Competition* for each campaign keyword. We use these parameters for observing and predicting the campaign behavior in favor of the advertiser.

Our main contributions are the following:

— An automated method for budget optimization based on a MCKP (multiple-choice knapsack problem) modeling and capitalizing on genetic algorithms to maximize profit or traffic, the two usual objectives for websites.
— A fully implemented and functional prototype system, developed for Google AdWords platform, which currently occupies a vast share of web-search advertising volume.
— A comprehensive experimental evaluation on real world data. Experimental results show that automated campaigns overall outperform the manual competitive ones.

---

Overall the proposed framework can contribute to considerably optimizing the resources (time and experienced personnel) devoted to developing and monitoring a campaign. On top of this the monitoring module ensures the maximization of the profit respecting the available budget.

This paper is organised as follows. In Section 2 we offer an overview of related work regarding challenges and key issues in Online Advertising. The description of the first component regarding Keyword and Ad Creative Generation can be found in Section 3 as well as a brief evaluation. In Section 4 we discuss and analyze the proposed strategy for Campaign Management and Optimization. An overall presentation of the integrated system is described in Section 5. Various case scenarios for experimental evaluation along with their results are presented in Section 6. Finally, in Section 7 we conclude and summarize the major points of the proposed framework as well as suggest future directions for expanding the functions of the prototype.

## 2. RELATED WORK

The research areas of sponsored search, textual advertising and keyword research involve among other topics the automatic extraction, suggestion, and expansion of keywords. An advertising campaign might involve more than one and in most cases a large number of landing pages. The manual selection of even a small set of keywords for advertising purposes is quite laborious, a fact which leads to the recent appearance of commercial tools that create keyword sets directly from a landing page. There exist different techniques for keyword generation. Search engines use query log based mining tools to generate keyword suggestions. In this way, they focus on discovering co-occurrence relationships between terms and suggest similar keywords. They start from an initial key phrase and they are based on past queries that contain these search terms. Google AdWords Keyword Tool [6] exploits this ability and presents frequent queries for the seed set of terms

Other commercial tools [7] determine an advertiser's top competitors and then actively search for the keywords they are targeting. After a period of time, lists of targeted keywords that are competitive for pay per click advertising are automatically generated. These two approaches may result to a recommendation set of keywords which are likely to be general and thus more expensive. Considering this, the challenge of generating keywords is to select both semantically similar and well-focused keywords.

TermsNet and Wordy [Joshi and Motwani 2006; Abhishek and Hosanagar 2007] exploit the power of search engines to generate a huge portfolio of terms and to establish the relevance between them. After selecting the most salient terms of the advertiser's web page they query search engines with each initial seed term. With their methods they find other semantically similar terms. Wordy system proposed single word terms (unigrams) for each seed keyword. S. Ravi et al. [Ravi et al. 2010] propose a generative model within a machine translation framework so the system translates any given landing page into relevant bid phrases. They first construct a parallel corpus from a given set of bid phrases $b$, aligned to landing page keywords $l$, and then learn the translation model to estimate $Pr(l|b)$ for unseen $(b, l)$ pairs. This approach performs very efficiently but depends on the chosen domain and data that the human decision factor may affect.

In general, *corpus or domain dependent* systems require a large stack of documents and predetermined keywords to build a prediction model [Liu et al. 2010], while on the other hand our developed system works with a *corpus independent* approach that directly sifts keywords from a single document without any previous or background information.

---

[6]http://www.adwords.google.com/keywordtool

[7]http://www.adgooroo.com/, http://www.wordstream.com/

Regarding the automated ad creative generation process, to the best of our knowledge, this issue remains still an open problem in natural language processing and information retrieval areas as mentioned in[Gabrilovich 2011]. Thus, the corresponding module of our system is an innovative contribution in this regard.

Assuming the ad auction mechanism of a search engine, the main issue that the advertisers are facing is to decide their bidding strategy and how they are going to split their budget among the keywords of their campaign. There have been various attempts to solve the budget optimization problem, some of which are based on heuristics, some calculate approximations using linear programming variations, and others take a more statistical and stochastic approach. Even Dar et al. [Even Dar et al. 2009] present their approach of maximizing profit, using a linear programming (LP) based polynomial-time algorithm. To deal with the NP-hardness of the problem, they propose a constant-factor approximation when the optimal profit significantly exceeds the cost. It is based on rounding a natural LP formulation of the problem. Szymanski and Lee [Szymanski 2006] discuss how advertisers, by considering minimum return on investment (ROI), change their bidding and, consequently the auctioneer's revenue in sponsored search advertisement auctions.

Borgs et al. [Borgs et al. 2007] propose a bidding heuristic that is based on equalizing the marginal ROI across all keywords, so they change each keyword bid based on the ROI performance of the previous day. Their system converges to its market equilibrium in the case of the first price mechanism with a single slot when everybody adopts the proposed perturbed bid solution.

Rusmevichientong and Williamson [Rusmevichientong and Williamson 2006] develop an adaptive algorithm that learns the proportions of clicks for different keywords by bidding on different prefix solutions, and eventually converges to near-optimal profits, assuming that various parameters are concentrated around their means. Muthukrishnan et al. [Muthukrishnan et al. 2010] consider stochastic algorithms that attempt to solve the problem in advance, and not by adaptive learning as in [Rusmevichientong and Williamson 2006], and work for pre-specified probability distributions of keyword clicks.

Zhou et al. [Zhou et al. 2008] model the problem of advertisers winning an ad slot for one keyword they bid upon as an online multiple-choice knapsack problem. Zhou and Naroditskiy [Zhou and Naroditskiy 2008] continue the work of [Zhou et al. 2008] modeling budget-constrained keyword bidding as a stochastic multiple-choice knapsack problem. Their algorithm selects keywords based on a threshold function which can be built and updated using historical data. It employs distributional information about prices and tries to solve the bidding problem with multiple ad-position, keywords, and time periods.

The problem of finding a near-optimal bidding strategy has been also approached by using autonomous agents. The TAC Ad Auctions (TAC/AA) game investigates complex strategic issues found in real sponsored search auctions through a simulation of the general auction process [Jordan and Wellman 2010].

## 3. GENERATING KEYWORDS AND AD-TEXTS

### 3.1. Keyword Generation Component

This component aims at proposing valid and representative keywords for a landing page capitalizing on keyword extraction methods, on the co-occurrence of terms, and on keyword suggestions extracted from relevant search result snippets.

Table I. Tag Weights

| Element | Assigned Weight |
|---|---|
| <title> | 50 |
| meta keywords | 40 |
| meta description | 40 |
| anchor text | 30 |
| <h1> | 30 |
| <b> | 10 |
| other | 1 |

*3.1.1. Keyword Extraction Module.* In this process, we follow the corpus independent approach to rely solely on the given landing page document. In a web page structure, the text fields represent the semantics of the page. According to vector space model, each web page can be considered as a document and its text content must be segmented as many weighted keywords which all together represent the semantics of a document [Zhou et al. 2007]. After segmentation of text, the result will be a set of keywords usually called a "term" in a document. Then each of these terms must be weighted properly to assure that terms with higher semantic meaning and relevance to our page have larger weight. As a preprocessing step, the HTML content of each landing page is parsed, stopwords are removed and the text content is tokenized. For this process, our system uses the Jericho HTML Parser [8], a Java library allowing for analysis and manipulation of parts of an HTML document, including server-side tags.

Next, for each word (gram) $l_j$ in the tokenized output, we compute a weight associated with the gram for each occurrence inside a specific tag, e.g. the occurrence of a gram inside $<h1>$ tags.

$$w_{jtag} = weight_{tag} * f_{jtag} \tag{1}$$

where $weight_{tag}$ is a special weight assigned to each different HTML tag and $f_{jtag}$ is the frequency of the gram inside the specified tag. The weight of each tag is assigned according to its importance inside the HTML document. We set higher values on important tags such as $<title>$, meta keywords, meta description, anchor text, $<h1>$, $<b>$. In Table I we propose the assignment of tag weights following an approach that ranks the importance of these tags according to where web page designers choose to place the most important information on their website. Then, we compute the special weight of each gram as the sum of all $w_{jtag}$ weights:

$$special\_weight_j = \sum w_{jtag} \tag{2}$$

In the next step, the relevance score of each gram is computed:

$$relevance\_score_j = \frac{special\_weight_j}{MAX\_WEIGHT} \tag{3}$$

where $MAX\_WEIGHT$ represents the maximum special weight that a gram could have inside the HTML document. $MAX\_WEIGHT$ can be different for each HTML document because some of them may not have links or bold tags, etc.

Unimportant unigrams occurring on the page are filtered out using a threshold $\tau = 0.001 * relevance_{max}$, resulting after several experiments and parameter tuning on the relevance score. While unigrams frequently have a broad meaning, multiword phrases (n-grams) are more specific and thus can be more representative as advertising keywords. A typical query length, especially while searching for a product or service, varies between 1 and 3 grams. For that reason, from the extracted single word terms (unigrams) we pull together possible combinations of two-word phrases (bigrams) inside the given landing page. Next, in order to construct the gram co-occurrence matrix, the top N grams with high relevance scores are ranked in descending order. Then we define co-occurrence as follows: if $gram_i$ and $gram_j$ appear in a same unit (each different area inside an HTML document, defined by HTML tags) which is predefined, then they co-occur once, and $freq_{i,j}$ should be increased by one.

---

[8]http://jericho.htmlparser.net/docs/index.html

Finally, we consider the co-occurring two-word terms (bigrams) above $\tau$ and follow the same process, searching for new co-occurrence with each unique unigram. In this way, we extract three-word terms (trigrams) as well. By gathering all terms, we construct the extracted keywords vector. In order to boost trigrams first, bigrams second and unigrams third, we modify their relevance score with the following factor:

$$boosted\_score_j = relevance\_score_j * k^{noOfGrams} \qquad (4)$$

where $k$ is a free parameter (in our experiments we set it to ($k = 100$) and $noOfGrams$ is the number of grams composing a term.

*3.1.2. Keyword Suggestion Module.* From the previous step of keyword extraction we have already extracted the initial keywords. These will be the seed keywords for the additional suggestions. Initially, as this procedure begins, the set of additional suggestions is empty. We provide as input the extracted keywords from the landing page. For each given seed keyword, the keyword is submitted as a query $q$ into a search engine API. We use for this purpose Google JSON/Atom Custom Search API [9]. With this API, developers can use RESTful requests to get search results in either JSON or Atom format. The API returns a set of short text snippets, snippets that are relevant to the query and thus to the keyword.

From the response data we retrieve `feed/entry/summary/` `text()` which is a string type property indicating the snippet of the search result and `feed/entry/title/text()` which is a string type property indicating the title of the search result. The top 30 results are downloaded and loaded in Apache Lucene Library [10], which we use it for implementing indexing and query support in our system. Each extracted term from the previous step which was a seed for the query has now been extended to a set of results which we use as a document in the Lucene index. Each set of title and snippet results that were retrieved after a seed query represents this document $d$ for Lucene indexing.

In this step, we parse the resulting document and construct a new vector of grams $< g_1, g_2, \ldots, g_{|d|} >$. Based on the Lucene scoring method we find the unigrams and bigrams that have the largent number of occurrences inside the document and that are kept as the most relevant for the specific seed query. Each of these terms is representing a new query $q'$.

The score of query $q'$ for document $d$ is considered to be the cosine-distance or dot-product between document and query vectors in a Vector Space Model (VSM) of Information Retrieval. Again, we sort in descent order the new queries based on this score and we create a vector of suggested keywords and their scores for each of the seed terms. Before we place our output as an integrated input vector to the next component, we normalize scores using min-max normalization:

$$relevance' = \frac{relevance_i - relevance_{min}}{relevance_{max} - relevance_{min}} \qquad (5)$$

---

[9]`http://code.google.com/apis/customsearch/v1/overview.html`
[10]`http://lucene.apache.org/java/docs/index.html`

where $relevance_{max} - relevance_{min} \neq 0$.

Finally, we use a new threshold ($\tau = 0.5$ which represents the 50% of the maximum relevance) for keeping only the most salient terms.

### 3.2. Ad Creative Generation Component

In an advertising campaign, the text-based advertisement comprises of a brief title and description, typically no more than 25 and 70 characters long and is associated with each keyword in the relevant ad groups. When a user submits a query to a search engine for a keyword appearing on the advertiser's list, the corresponding advertisement is displayed next to the organic search results, an action called "impression". The advertiser pays only when the ad is clicked, following the pay-per-click model. A typical sponsored-search campaign contains multiple bid terms, ranging from a handful to millions. Particularly for large campaigns, managing the copy of sponsored-search ads, their titles and descriptions presents a substantial editorial challenge if the task is going to be implemented manually. One option is to use the same title and description for every bid term. But industry wisdom holds this to be a poor strategy, because ads appear less targeted when they do not include the term itself [Bartz et al. 2008]. Both Google and Yahoo render instances of the term in bold text, so an ad with generic text suffers next to one with targeted copy. The alternative is to write a separate title and description for each term, but this can be a daunting task for large campaigns.

For this purpose, we propose an automated process for the *ad creative generation*. Summaries of Web sites help Web users get an idea of the site contents without having to spend time browsing the sites. The technology of automatic text summarization is maturing and may provide a solution to the information overload problem. Automatic text summarization produces a concise summary of source documents. The summary can either be a *generic summary* (this type of summary we using in the next process), which shows the main topics and key contents covered in the source text, or a *query-relevant* summary (which is a further challenge of our system as we could use specific keywords for filtering the resulted summaries), which locates the contents pertinent to user's seeking goals [Zhang et al. 2004].

Concretely, in this subprocess the first step was to extract all the text from the HTML document of the given landing page. Then, we used summarization to keep the most important meaning for the description of our advertising page. For this purpose the input was the text from the page to the Classifier4J [11] which is a Java library designed to do text classification. It comes with an implementation of a Bayesian classifier, and has also some other important features, including a text summary facility.

The features for a Google AdWords text ad are the following:

(1) Headline of the textAd $head$: The problem or opportunity; Ad titles are limited to 25 characters
(2) Description Line 1 $dl_1$: Short description of big benefit; limited to 35 characters
(3) Description Line 2 $dl_2$: Short description of the product/service; limited to 35 characters
(4) DisplayURL $url_{display}$: The web site's name up to 35 characters; Google can only display up to 35 characters of the display URL, due to limited space. If the display URL is longer than 35 characters, it will appear shortened when the ad is displayed
(5) DestinationURL $url_{destination}$: Landing page: URL of the exact Web page customers visit first

We kept the constraints and limitations that are given from Google AdWords platform (number of characters in the ad lines) and we added at the end of the second description line a call-to-action phrase such as: "Buy now!", "Purchase now!", "Order now!", "Browse now!", "Be informed" according to each advertising goal. For example we use "Be informed" for advertising a web page or a brand name, "Buy now!" for advertising a product, "Purchase now!" for advertising a service. The purpose of adding these phrases into the text ad is to optimize the ad descriptions because a call-to-action encourages users to click on the ad and ensures they understand exactly what the advertiser expects them to do when they reach the landing page. We present in Algorithm 1 the procedure of the ad-text automatic creation and in Figure 1 a demonstration of the output.

---

[11]http://classifier4j.sourceforge.net/

---

**ALGORITHM 1:** Ad-text automatic creation

---

**Input**: Landing Page HTLM Document $d$, $t$ the target of the advertisement
**Output**: An ad-text
Let $\lambda$ be the length of a sentence in characters
Let $\phi$ be the set of the action phrases
$limit_1 = 25, limit_2 = 35$
▷ Choose a proper action phrase $p_{action} \in \phi$ with respect to $t$
$p_{action} \mapsto t$
$bidPhrase \leftarrow keywordGenModule(url_{destination})$
▷ Retrieve the first phrase of the title until the first punctuation
$title = < p_1, p_2, \ldots, p_n >$
**if** $\lambda_{p_1} < limit_1$ **then**
        $head \leftarrow p_1 \cap bidPhrase$
**else**
        $head \leftarrow bidPhrase$
$head \leftarrow capitalizeFirstLetterOfGrams(head)$
▷ Summarise $d$ in 1 sentence using Bayesian classifier
$d_{summary} \leftarrow summariser(d, 1)$
$d_{summary} = < s_1, s_2, \ldots, s_N >$
**while** $\lambda_{\bigcap_{i \in N} s_i} \leq limit_2$ **do**
        $dl_1 \leftarrow \bigcap_{i \in N} s_i$
**end**
**while** $\lambda_{(\bigcap_{k \in N} s_k) \cap p_{action}} \leq limit_2$ **do**
        $dl_2 \leftarrow \bigcap_{k \in N} s_k$
**end**
**if** $s_{final}$ *is a stopword* **then**
        remove $s_{final}$ from $dl_2$
$dl_2 \leftarrow (dl_2 \cap p_{action})$
$url_{display} \leftarrow "www." \cap p_1 \cap ".com"$
$adText \leftarrow (head \cap dl_1 \cap dl_2 \cap url_{display})$
**return** $adText$

---

### 3.3. Experimental Evaluation of GrammAds Component

In our initial experiments described in [Thomaidou and Vazirgiannis 2011], we evaluated at a first glance our method using human ranking for resulted keywords following a blind testing protocol. The landing pages for our experiments were taken from different thematic areas, promoting several products and services. The categories were: {*hardware product, corporate web presence optimization service, gifts, GPS review, hair products, vacation packages, web design templates, car rental services*}. In

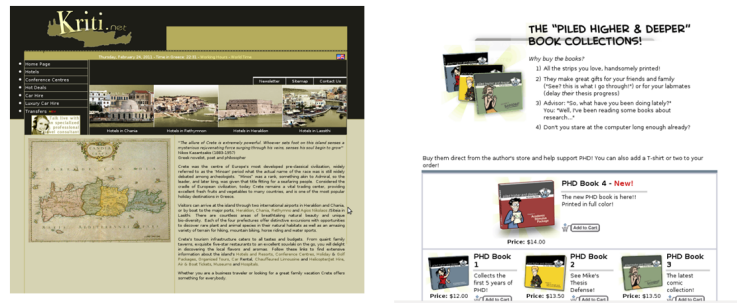

Crete Vacation Packages
The allure of Crete is extremely powerful. Book now!
www.cretehotels.com

Piled Higher Full Cover
Buy them direct from the author's store and help support Buy now!
www.piledhigheranddeeper.com

Fig. 1.    Automated Generated Ad Creatives

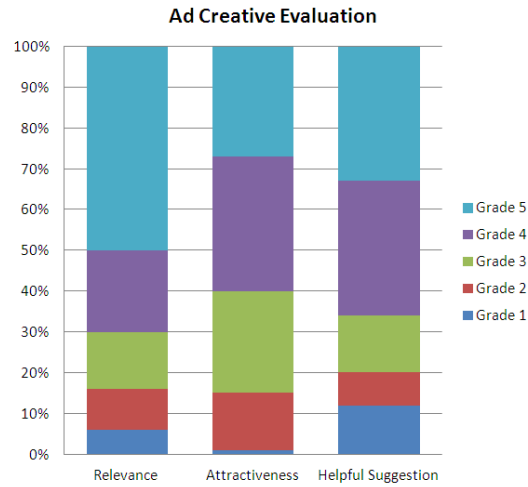

Fig. 2.    Ad Creatives Feedback

the next steps we evaluated also the automatically generated ad-texts. Eleven researchers and postgraduate students provided feedback in the scale of Grade.1:Bad up to Grade.5:Very Good. In Figure 2 we present the evaluation results. In the context of various experiments where we used the exact same bidding strategy for two identical campaigns of a company that offers web developing solutions (a highly competitive field for online advertising), we discovered the following: The keywords that were generated form GrammAds *achieved higher Clickthrough rate (CTR) values* than the manual inserted ones as shown in Figure 3

## 4. BUDGET OPTIMIZATION PROBLEM
The most challenging issue in the managing process of an advertising campaign is the *Budget Optimization* for the multiple keywords of the campaign. We consider the

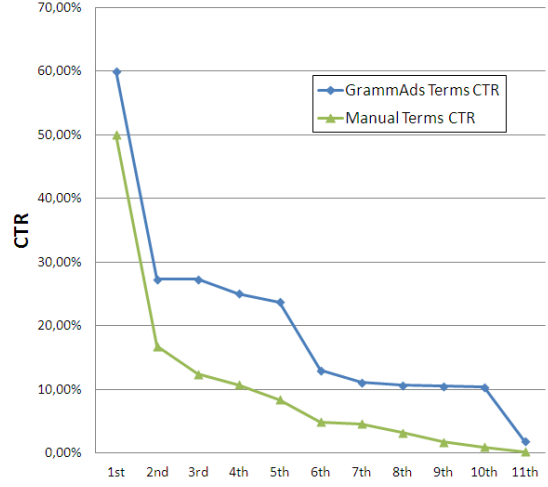

Fig. 3. Keyword CTR Comparison for top-11 generated terms

problem as follows: Assuming a limited budget $B$, we aim to find the combination of keywords with bids that maximizes the campaign profit. In particular, we are looking for a set of keywords $k \in K$ ($K$ is the set of all possible relevant keywords), and their bids $b \in \mathbb{R}_{\geq 0}$ with

$$\sum_{k \in K} w_k(k, b) \leq B \qquad (6)$$

where $w_k$ is the bidding cost $b$ on keyword $k$ (otherwise called weight) that produce:

$$\max \sum_{k \in K} v_k(k, b) \qquad (7)$$

where $v_k$ is the function that computes the expected profit of keyword $k$ (value) assuming of bidding price $b$. We also consider that for any given $k$ with $b = 0$, $b_k = 0 \Rightarrow w_k(k, b) = 0$ and $v_k(k, b) = 0$. A zero bid actually means that we choose not to bid on the particular keyword, so there is no cost or profit produced. In the following sections, we present our approach of finding the best combination of keywords and bids that produce maximum profit. Profit can be either monetary profit from product sales or generated traffic (clicks on ads) for the advertiser's website. We define the above concepts as follows:

DEFINITION 1. (Weight and Cost) *The cost of a keyword $k$ for a given bid $b$ is the product of expected number of clicks and the average cost per click.*

$$w(k, b) = \overline{CPC}(k, b) * Clicks(k, b) \qquad (8)$$

In Definition 1 $Clicks(k, b) = CTR(k, b) * Impr(k, b)$, $CTR$ = Click-through-rate, $Impr$ = Impressions, $\overline{CPC}$ = Average-cost-per-click

DEFINITION 2. (Value for maximum monetary profit) *The profit from each keyword-bid combination comes from subtracting the cost of clicks, which is the cost of advertisement, from the revenue of sales.*

$$v(k, b) = Revenue(k) * CR(k, b) * Clicks(k, b) - w(k, b) \qquad (9)$$

In Definition 2 $CR(k, b) * Clicks(k, b)$ is the total conversions (sales) that we expect to have and $Revenue(k) * CR(k, b) * Clicks(k, b)$ is the revenue expected for $(k, b)$, $CR$ = Conversion-rate, $Revenue$ = Revenue-per-conversion.

DEFINITION 3. (Value for maximum traffic) *When we are interested in maximizing the traffic led to a website, the only valuable measure is the amount of clicks that are generated from keywords.*

$$v(k, b) = Clicks(k, b) \tag{10}$$

### 4.1. Multiple-choice knapsack problem formulation

In the online advertising campaign, the advertiser plays the role of an investor. The capital is the total budget $B$ for the period that the campaign is active. The profit from the conversions or clicks for each investment is represented as $v$. The cost that the advertiser is finally charged for a specific investment is $w$ . Each investment is represented by a candidate item $x$ which is a pair $(k, b)$ where $k$ is the keyword and $b$ the bid that the advertiser initially sets as maximum $CPC$ for the specific keyword. The advertiser has $j$ options of $(k, b)$ candidate pairs for each investment, but he must select only one pair per investment for his final proposal, because for a particular keyword $k$ in the auction process, he can set only one bid. The total number $N$ of the final chosen investments must be equal to the $r$ available keywords of the campaign. This is a Multiple-Choice Knapsack Problem (MCKP). MCKP is a 0-1 knapsack problem in which a partition $N_1 \cdots N_r$ of the item set $N$ is given and it is required that exactly one item per subset is selected. Formally for our problem, the objective is to

$$\text{maximize} \quad \sum_{i=1}^{r} \sum_{j \in N_i} v_{ij} x_{ij}$$
$$\text{subject to} \quad \sum_{i=1}^{r} \sum_{j \in N_i} w_{ij} x_{ij} \leq B \tag{11}$$

$$\text{with} \sum_{j \in N_i} x_{ij} = 1 \text{, for all } 1 \leq i \leq r \tag{12}$$
$$\text{and } x_{ij} \in \{0, 1\}\text{, for all } 1 \leq i \leq r \text{ and all } j \in N_i$$

The above imply that only one bid option is going to be selected for each keyword.

The optimal solution of the MCKP will indicate the best possible choice of keyword-bid options. Our approach was to model this combinatorial optimization problem in a certain way where we can also formulate it as a genetic algorithm (GA) process. In MCKP, the goal is to find for each keyword the option that maximizes the achieved profit. In GA, different chromosomes represent different instances of candidate items and the goal is to find the fittest chromosomes. As we will describe later, a GA finds approximately the proper options of MCKP for profit maximization. This process aims to collect proper statistics from previous time periods and keep only the most profitable options for the next time period. This problem formulation, as we can see in Figure 4, is different from the approach that we have seen in [Zhou and Naroditskiy 2008] and [Zhou et al. 2008], as in our method we focus on the clicks each keyword gains, rather than use MCKP to model the ad auction policy, where each advertiser can select to win at most one ad slot for each keyword.

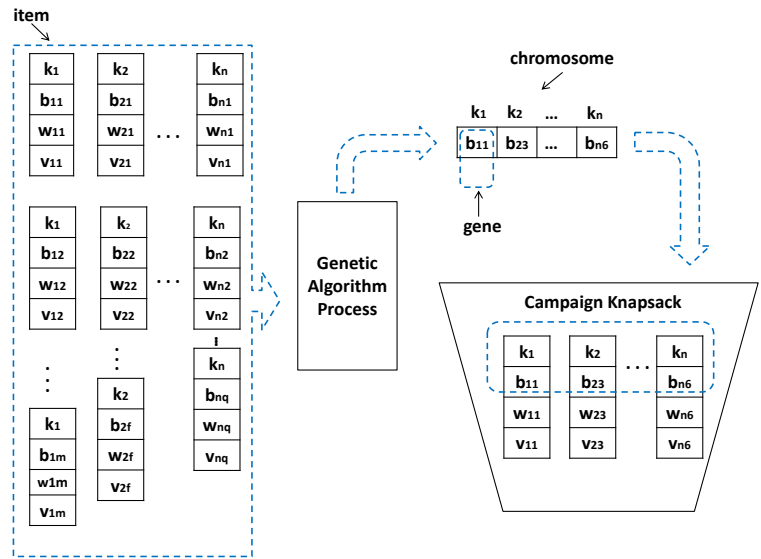

Fig. 4. Mapping of campaign system to the MCKP. Items: options of keyword-bid pairs along with their profit v and cost w. Chromosome ≡ Set of selected items

## 4.2. Genetic Algorithm Advantages

Multiple-choice knapsack [Martello and Toth 1990] is a known NP-Complete problem, although some solutions for approximate optima in (pseudo-)polynomial time have been found. The approach we take is to capitalize on genetic algorithms [Djannaty and Doostdar ] that have also polynomial complexity and are used in a variety of global optimization [Mitchell 1996], [Weise 2008] problems. GAs find optima in certain search spaces and are able to combine *exploration*, the process of discovering possible solutions in search spaces, and *exploitation*, the process of using the knowledge of past solutions (past generations) to the benefit of a new more advanced solution. GA finds optimal solutions or near-optimum, since it is an approximation method, like any other polynomial time method that exists. Deterministic methods result in the same approximate solution in each run, thus making it difficult to collect data for many keywords. This is because this method will use those keywords that were chosen repeatedly in the past.On the other hand, the solution of a genetic algorithm may vary, resulting in a different near-optimum solution in each run. This trait is an advantage, as we do not want our method to have obsessions with certain solutions, thus choosing persistently certain keywords. This kind of flexibility, allows our system to discover faster whether keywords are performing better or worse than they did in the past. Therefore, as ad campaign parameters change, something frequent in the case of ad auctions, a deterministic method will adapt much slower than a genetic algorithm.

## 4.3. Bidding Strategy

The objective is to create a population of candidate solutions (called chromosomes). In each successive generation, a new population of chromosomes is produced by combining (with a procedure called crossover) pairs of chromosomes of the last generation to create new chromosomes (reproduction). The best chromosomes have a better chance to reproduce in the next generation (survival of the fittest), ensuring that each generation is improving. *Selection* is the process of finding the fittest chromosomes to become the parents of the next generation. For this purpose, there are fitness-proportionate techniques such as *Weighted Roulette Wheel Selection* (Weighted RWS) and *Stochastic Universal Sampling* (SUS). These methods make sure that, if a chromosome has a strong fitness, it will have proportionately high probability of reproducing. Moreover, we make sure that a (small) proportion of the fittest chromosomes pass directly to the next generation. This action is called *elitism* and its purpose is to prevent loosing the few best found solutions, increasing the performance of the genetic algorithm. The process of combining two chromosomes is called *crossover*. Every time, two offsprings are produced by two parents and the parents are replaced. The first offspring takes a part of each parent while the other obtains the remaining part of the parents. We want our genetic algorithm to avoid falling in local optima. Thus, the concept of mutation is applied on the chromosomes after the crossover process. Mutation changes the new offspring by altering, with a small probability, the value of their genes increasing the chance for reaching to the global optimum. The process of generating new populations terminates, usually, when $\sim 90\%$ of the chromosomes have the same fitness value or the highest ranking solution's fitness has reached a plateau, i.e. successive iterations no longer produce better results. Alternatively termination occurs in the case number of generations is greater than a certain limit.

### 4.3.1. Strategy Steps.

In this part, we present the main process of our methodology implementation. First, we describe the primary steps for initializing system parameters. We must define a default initial bid for all keywords that are going to be tested, so given a specific variable information from Google AdWords, we set $b_{initial} \leftarrow maxEstimatedFirstPageBid$. Next, we define time for task periods (e.g. 2 days) and adgroups for each landing page along with their keywords and text ads. In Algorithm 2, we describe the general form of training periods to test campaign and adgroup settings in order to collect proper statistics. The genetic algorithm step is the implementation of the optimization process. Finally, in each testing phase after optimization, we follow the same process of the first training periods but we also pause previous keywords that are not selected by the optimization module.

The reason for different training periods with only a small amount of testing keywords maintained is that, due to a limited daily budget for our experiments, we do not want to exhaust the budget without having tested many keyword options.

### Optimization step

In the genetic algorithm, the bids for each keyword that are available to choose from are ones that have been tested and we have kept statistics on. It is not possible to have full information about the performance of a keyword that has not been tested at some point. Important performance criteria for our method are *click-through rate*, *impressions*, *average cost-per-click* and *conversion-rate*. Once we have collected the proper statistics, we are ready to apply our genetic algorithm for optimization.

### Genetic Algorithm Representation

*1. Start.* Generate random population of $m$ ($m = 40$) chromosomes
**Chromosome Representation**
— For the budget optimization problem, each chromosome consists of N genes, N being the number of available keywords

  Each gene has a value of the bid index that is selected for the specific keyword
— Table II shows a chromosome that has selected the second bid for keyword $k_1$ and zero bid (value 0) for keyword $k_2$

---

**ALGORITHM 2:** Training Period

---

**Input**: Settings of Adgroups
**Output**: Collected statistics
Let $t$ be the number of task periods
Let $S_G \subset N_G$, where $N_G$ are all the candidate keywords of AdGroup $G$
▷ Make a subset of $|S|$ keywords for testing for each AdGroup $G$
▷ $|G|$ is the total number of AdGroups
$|S_G| \leftarrow |N_G|/t$
**forall the** $g \in |G|$ **do**
    add(AdGroup[g].getMostRelevantKeywords($|S_G|$), keyword)
**end**
$|M| \leftarrow |G| * |S|$
**forall the** $\mu \in |M|$ **do**
    setBid($b_{initial}$, keyword[$\mu$])
    activate(keyword[$\mu$])
**end**
**if not** $firstPeriod$ **then**
    chooseRandom(keyword)
    **forall the** $\mu \in |M|$ **do**
        **if** $choosed(keyword[\mu])$ **then**
            $bidNew[\mu] = bidPrevious[\mu] \pm bidPrevious[\mu] * 50\%$
        **end**
    **forall the** $\mu \in |M|$ **do**
        ▷ Do not test again other keywords that received clicks
        **if** $notChoosed(keyword[\mu])$ **and** $receivedClicks(keyword[\mu])$ **then**
            pause(keyword[$\mu$])
    **end**
**while** $taskPeriod > 0$ **do**
    **forall the** $\mu \in |M|$ **do**
        stat[$\mu$] = collect(impressions[$\mu$] $\cap$ clicks[$\mu$] $\cap$ conversions[$\mu$] $\cap$ averageCPC[$\mu$])
        add(stat[$\mu$], statistics)
    **end**
    $taskPeriod \leftarrow taskPeriod - 1$
**end**
**return** $statistics$

---

— Table III shows that the second bid ($bidIndex = 2$) for $k_1$ is the actual bid value of $0.60
— Table IV shows that this bid has a cost of $16.2 and a positive profit of $1.40. If a keyword is not selected ($bidIndex = 0$), like $k_2$ in Table II, it produces zero cost and profit

*2. Fitness.* Fitness Function Evaluation
— The fitness function is the expected total profit for the bids selected in the chromosome genes.

$$\text{Chromosome Fitness} = \sum v(k_i, b_i) \tag{13}$$

Fitness function resembles the objective function of the knapsack problem. It can be easily computed since we have pre-computed all the costs and profits of the bids for every keyword, as shown in Table IV

Evaluate the fitness function of each chromosome in the population. Take into consideration *actual* or *predicted* values

— When a chromosome is generated, it has to pass the $\sum w(k_i, b_i) \leq B$ condition, otherwise randomly selected genes of the chromosome will be set to 0 until the condition is met

*3. New Population.* Create a new population by repeating the following steps until the new population is complete:

*a. Selection.* Select two parent chromosomes from a population according to their fitness (Weighted RWS). The best chromosomes are the ones with the highest values of the fitness function

*b. Crossover.* With a crossover probability, cross over the parents to form a new offspring (children). If no crossover was performed, offspring is an exact copy of parents

*c. Mutation.* With a mutation probability ($\sim 0.1\%$) mutate new offspring at each locus (position in chromosome)

*d. Accepting.* Place new offspring in a new population

*4. Replace.* Use new generated population for a further run of algorithm

*5. Test.* End Condition

— Since we don't know what the best answer is going to be, we just evolve the max number of times *(MaxAllowedEvolutions = 3000)*

If the end condition is satisfied, **stop**, and return the best solution in current population

— *Loop.* Go to step **2**

After a period of testing and collecting statistics, budget optimization task is ready to run again. This process continues executing until the last day of our campaign.

*4.3.2. Impressions Prediction.*

We wanted also to examine if a certain type of campaign behavior prediction could be helpful to our system. Clicks, click-through rate, and conversion rate are parameters that are more dependent on inner factors of the advertiser's choices such as the quality and relevance of the selected keywords and ad-texts for the product promotion. However, this is not exactly the case for the impressions that a user query generates. The impressions fluctuate primarily and because of other factors external to the keyword-bid combination. Consequently, we need a means to predict or at least make a good estimation of how many impressions a keyword will receive matched with a specific bid, knowing:

(1) Past received clicks for various selected $(k, b)$ combinations
(2) Current average user searches for a query similar to this keyword
(3) Current will of competition of all the other bidders upon this specific keyword

The idea is to use past results of keyword behavior in a model that can capture externalities of the ad auctions and predict current or future behavior. Google AdWords provides information such as Global Monthly Searches (GMS) and Competition of a keyword which are factors that affect the number of impressions of a keyword and, at the same time, are independent of a particular AdWords Account.

Past data of all keywords with known Impressions has the following form:

$[Clicks(k_1, b_1), GMS(k_1), Competition(k_1)] \rightarrow Impressions(k_1, b_1)$
$[Clicks(k_2, b_2), GMS(k_2), Competition(k_2)] \rightarrow Impressions(k_2, b_2)$
$\ldots \qquad\qquad\qquad\qquad\qquad \ldots$
$[Clicks(k_n, b_n), GMS(k_n), Competition(k_n)] \rightarrow Impressions(k_n, b_n)$

thus, we aim to predict the impressions of another keyword $-$ bid $(k_i, b_i)$:

$[Clicks(k_i, b_i), GMS(k_i), Competition(k_i)] \quad ? \rightarrow Impressions(k_i, b_i)$

Table II. Example of chromosome representation and
the values of its genes

|  | $k_1$ | $k_2$ | $k_3$ |  | $k_N$ |
|---|---|---|---|---|---|
| *bidIndex* | 2 | 0 | 3 | ... | 1 |
| *valueRange* | [0-4] | [0-3] | [0-3] |  | [0-2] |

Table III. Bid matrix example. For $k_1$, we have a value range
of [0-4] of the bidIndex

| bidIndex | $k_1$ | $k_2$ | $k_3$ | ... | $k_N$ |
|---|---|---|---|---|---|
| 1 | $ 0.50 | $ 0.90 | $ 0.45 | ... | $ 0.55 |
| 2 | $ 0.60 | $ 1.10 | $ 0.55 |  | $ 0.70 |
| 3 | $ 0.70 | $ 1.30 | $ 0.65 | ... |  |
| 4 | $ 0.80 |  |  |  |  |

Table IV. Example of expected costs and profits for each different $(k, b)$ pair

| $k_1$ | | | ... | $k_N$ | | |
|---|---|---|---|---|---|---|
| **bidIndex** | **w(k,b)** | **v(k,b)** | ... | **bidIndex** | **w(k,b)** | **v(k,b)** |
| 1 | $ 14.5 | $ 1.5 | ... | 1 | $ 11.0 | $ 1.5 |
| 2 | $ 16.2 | $ 1.4 | ... | 2 | $ 16.5 | $ 1.9 |
| 3 | $ 18.1 | $ 0.3 |  |  |  |  |
| 4 | $ 19.8 | $ 0.5 |  |  |  |  |

After prediction of impressions of all keyword-bid combinations is carried out, new values for clicks and conversions can be computed. A good estimation of impressions may result in a good cost and profit estimation, and can possibly lead to an improved budget optimization. To perform impressions prediction, we choose multiple linear regression [Montgomery et al. 2007]. This method assumes the existence of linear correlation between the dependent variable $y$ (Impressions) and the independent variables (in our case $x_1$ = Clicks, $x_2$ = GMS, $x_3$ = Competition). So, we need to find the best coefficients that show the relationship between $y$ and $x_i$. The goal is to be able to calculate a new value of $y$ out of the independent variables and the coefficients.

$$y' = \theta_0 + \theta_1 x_1 + \theta_2 x_2 + ... + \theta_k x_k \qquad (14)$$

In our case, we have 3 independent variables:

$$y' = \theta_0 + \theta_1 x_1 + \theta_2 x_2 + \theta_3 x_3 \qquad (15)$$

The regression model is fitted with the least squares [Lawson and Hanson 1995] approach. The sum of square residuals is considered to be the error ($e$) when comparing the $y$ with $y'$:

$$e = \sum_{i=1}^{N} \left(y_i - y'_i\right)^2 \qquad (16)$$

$N$ is the amount of all available records. In the end, the chosen coefficients $\theta_i$ must minimize the error produced by prediction to have the best fit of our model.

The result from this process is an alternative input to the genetic algorithm with different calculated statistics in order to study if using impressions prediction would achieve better campaign performance.

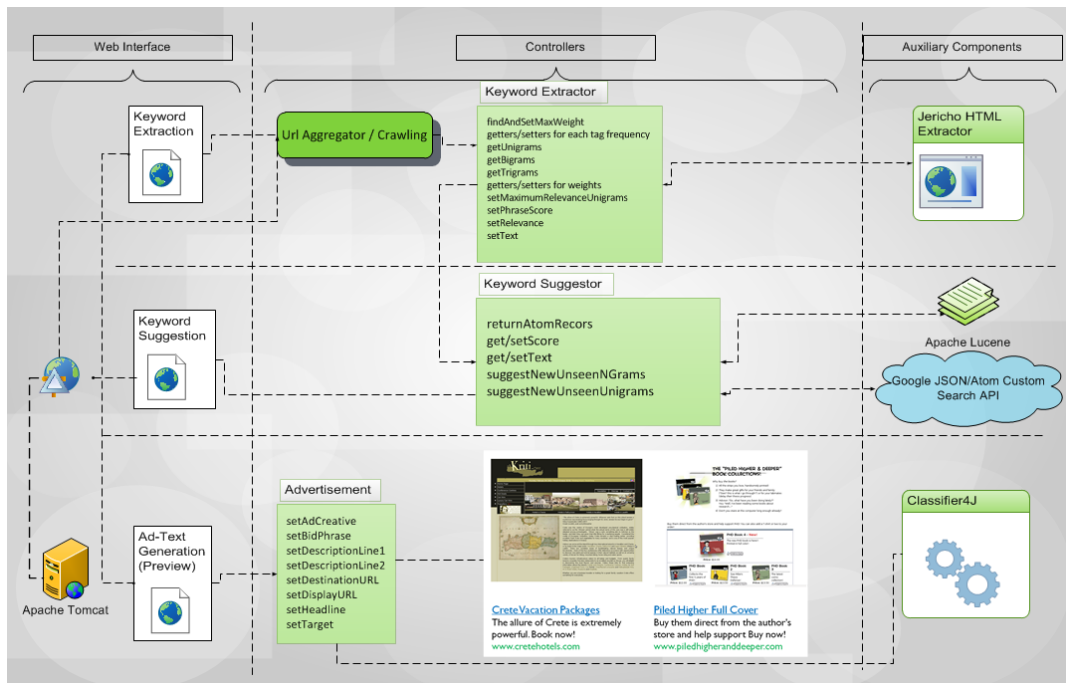

Fig. 5.   Initial Automated Creation

## 5. INTEGRATED SYSTEM FOR CAMPAIGN CREATION, MONITORING, AND OPTIMIZATION

### 5.1. Architecture

This system was developed in the context of an overall automated solution for creating, monitoring, and optimizing a Google AdWords campaign [Liakopoulos 2011; Thomaidou 2011].

The AD-MAD System that we have developed is separated in two parts:

(1) *The Keywords part (GrammAds Component)* as we present in Figure 5, which is responsible for retrieving the most relevant keywords and generating ad creatives based on information taken from the landing pages. [12] An initial approach of this component (regarding only the keyword selection part) is described in [Thomaidou and Vazirgiannis 2011]. The output of this part which generates multiword keywords (n-grams) and automated ad creative recommendations is selected as input feature in the following component.

(2) *The Campaign part (Adomaton Component)*, which is responsible for initializing, monitoring, and managing the advertising campaigns in the course of time, based on keyword statistics maintained by the system, in the view of optimizing available budget. An initial approach focused on the budget optimization process is described in [Liakopoulos et al. 2012].

In Figure 6, we present the two parts with the associations of their components.

The *Adomaton Component* is responsible not only for the proper monitoring and optimization of the campaign but also for the overall integration of the system. Once the right keywords are selected for each product and the ad texts are written, everything is ready to create and manage the advertising campaign, with the ultimate goal of increasing the web traffic or sales of the advertiser's website, according to the assigned preferences that the system user has given. In the next section, we will see the major tasks that the system must perform to reach the desired goal.

---

[12] A demonstration of the process can be found in the developed web application at www.grammads.com

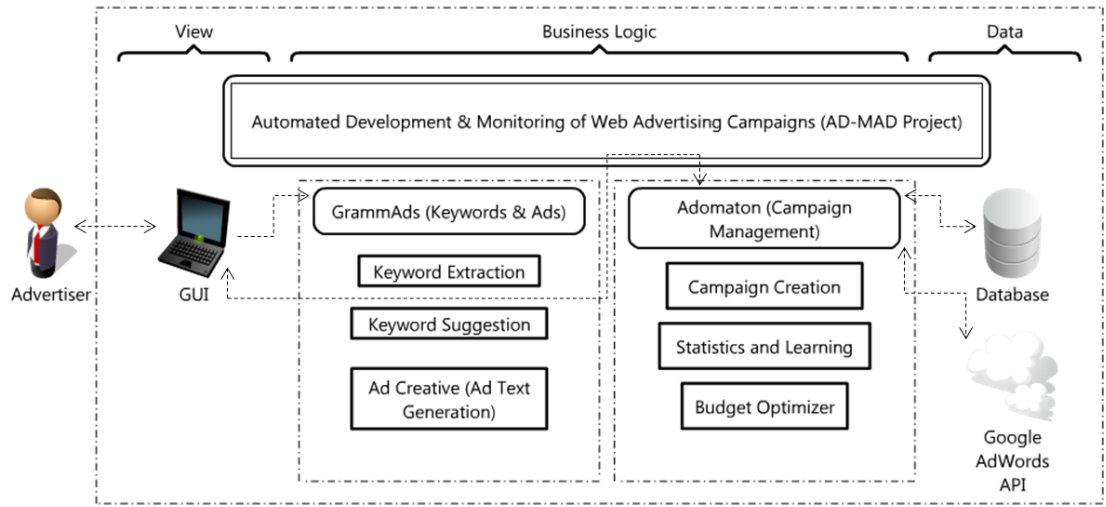

Fig. 6.   AD-MAD System - Components and their Modules

## 5.2. Software Description

The software that has been developed serves three basic purposes:

— To showcase that the proposed architecture and methodology is plausible and feasible and provide proof of concept.
— To become a good basis for a larger system that has more features and can provide more functionality useful for the field of web-search advertising.
— To find and implement tangible solutions to the issues arising at the very core of a Campaign creation and Budget optimization software.

The code of this software was completely written in Java using JDK SE 6 on the Eclipse IDE. For the database engine, MySQL 5.5 Community Server was used. We used AdWords API Java Library to communicate with the programming interface of AdWords, JGAP Library which is an open source framework for implementing and configuring genetic algorithms and genetic programming, and Flanagan's Java Scientific Library to implement multiple linear regression.

### 5.2.1. Campaign Creation.
#### AdWords API Wrapper

The Google AdWords API is fairly tricky. Even for trivial tasks, such as creating and setting a keyword, it requires a series of functions to be called, making it a long procedure to code. For the sake of programming sanity, many tasks must be wrapped to simpler functions, saving effort in typing, code readability and making the software less error prone. Furthermore, a lot of attention must be paid to the fact that using the AdWords API is not for free. In particular, Google charges API units per thousand and there is a rate-sheet [13] of every API-call. So, it makes sense to build a wrapper that balances usability with low-costliness.

*Create the campaigns*

The campaigns must be organized into adgroups which in turn contain keywords and ad creatives. Through the graphical user interface (GUI), it is requested from the advertiser to give, as input, the website of the promoted service. The Crawling module (i.e. a URL Aggregator) retrieves all the candidate landing pages of the inserted homepage. Using the extraction tools from the Keyword part of the system, those landing pages are analyzed and information is extracted, such as product information, languages, and type of advertisement needed. This information is verified and corrected by the advertiser if the system works in semi-automatic mode, otherwise the system uses this information as it is provided, working fully-automatically. Then, campaigns are created that contain adgroups and the rule of thumb is that for each landing page there must be a separate adgroup that contains its keywords and ad creatives. At the creation stage, all keywords work in trial mode, which means that no statistical data exist for them yet and the system waits to see how they will perform. After the first statistics are collected, the system can then use budget optimization.

### 5.2.2. Statistics and Database.

*Task Scheduler*

Since ad campaigns run for weeks, months, or even years, a scheduler is important for the system in order to ensure a task start and end. If a scheduler does not exist, the system cannot store proper statistics and allocate resources properly. The system also has to respect the budget constraint for a certain time period. For each new period, a new budget must be allocated and new optimal keywords must be selected.

*Statistical Information*

A SQL database must store all the statistical information collected for the keywords to be able to track their performance. All information about keywords, such as impressions, clicks, click-through rate, conversion rate, average position, average cost-per-click are necessary for the budget optimization module function. We also store temporal information, in the sense of distinguishing time periods. Also, there is the need for some data access objects (DAOs) that make retrieval and insertion of data to the database simpler. Both the Genetic Algorithm and the Prediction modules make extensive use of the statistical information of the keywords.

### 5.2.3. Budget Optimization.

*Genetic Algorithm*

The statistics module must provide a list of all possible keywords. Each keyword of the list must have all the possible bids for which we have statistics. For each keyword-bid combination $(k, b)$, an evaluation of the cost and profit must be computed based on the keyword performance information kept in the database. If a bid gives negative expected profit, it must be erased from the data because it cannot contribute to a maximum profit solution. Optionally, instead of directly using the value of keyword impressions to compute cost and profit, prediction can used. The predicted impressions will have an effect on expected clicks that in turn affect cost and profit. The problem is then modeled into chromosomes and the fittest chromosome is finally selected by the genetic algorithm after several generations of breeding using the methods of crossover, mutation, and elitism. A final list of keyword-bid pairs $(k, b)$ is produced from the fittest chromosome. These keyword-bid pairs will form the new bidding strategy, which must be set in the AdWords account of the advertiser using the AdWords API. Evolving the population the max number of times (we set the number of maximum allowed evolutions to 3000) and setting the initial population size of chromosomes to 40, help us avoid premature convergence, which was considered as the main problem of GA theory.

---

[13]http://code.google.com/apis/adwords/docs/ratesheet.html

*Prediction*

This is the module that is responsible for performing impressions prediction using the past statistics of keywords and targeted information taken from AdWords, such as the Global Monthly Searches and the level of Competition that exists for a given keyword. Before budget optimization, optionally, we can use this module in order to refine the statistics used by applying impressions prediction. A regression model finds the relationship between the impressions and other variables, such as clicks, global monthly searches, and competition. It then re-computes the value of impressions for each keyword-bid combination using the independent variables. The values of Global Monthly Searches and Competition must be computed from the AdWords API.

## 6. SYSTEM EVALUATION AND EXPERIMENTS

We present here several experiments and data analysis that we have conducted in order to study the performance of our proposed methodology and system. In the first three experiments, we evaluate only the optimization step, while in the fourth experiment we test the campaign creation, management, and optimization modules as an integrated process to evaluate the system as a whole.

### 6.1. Evaluation Data

We use the historical data of a large scale AdWords Campaign of a web site in the area of car rental. We selected from the collected data all the campaigns and adgroups that promote "car rental in Crete". The data collected are relevant to derive the period May 2009 to November 2010, during which the campaign was very active the majority of the time, generating traffic and sales for the car rental website. With the retrieved data of Google AdWords keyword statistics and sales for the car renting business, we get sufficient data for 39 weeks of this large scale campaign to perform tests on the impressions prediction and budget optimization modules. The final form of the integrated statistics table contains the following features: {*Campaign, Adgroup, Week, Keyword, MaxCPC, Impressions, Clicks, Conversions, CTR, Avg CPC, Cost, Profit, Quality Score, FirstPageCPC, Avg Position, Avg CPM*}. For the impressions prediction module of the system, we need to have for every keyword the "Global Monthly Searches" and the "Competition" values, retrieved using the AdWords API. Our budget optimization system provides two options; to optimize budget for maximum traffic or for maximum profit. Additionally, we can use original or predicted impressions.

Our budget optimization system provides two options; to optimize budget for maximum traffic or for maximum profit. Additionally, we can use original or predicted impressions. These options give us four basic testing scenarios:

(1) Budget Optimization for Profit with No Prediction (NoPredProfit)
(2) Budget Optimization for Traffic with No Prediction (NoPredTraffic)
(3) Budget Optimization for Profit With Prediction (PredProfit)
(4) Budget Optimization for Traffic With Prediction (PredTraffic)

### 6.2. Experiments

*6.2.1. Scenario Comparison.* The data used for the first experiment on the budget optimization process are the keyword statistics we collected from the car rental website for 39 weeks and the budget to be allocated for the next (hypothetical) week. Since budget optimization is performed with a genetic algorithm  a the stochastic method - the result will slightly vary every time it is executed, even with the same input data.

So, each scenario (NoPredProfit, NoPredTraffic, PredProfit, PredTraffic) is executed 30 times and the result reported is the average value of 30 executions.

The result of every execution of the budget optimization module is an optimal keyword-bid combination that ensures either maximum traffic or maximum profit for a limited budget. In particular, every result of the genetic algorithm application produces the following data:

— *Clicks*: How many clicks is the optimal solution (keyword - bid combination) expected to produce in the following week? This is an estimation, so it is represented with a double instead of an integer value
— *Cost*: How much is it expected to cost in the following week? This value must always be lower or equal to the budget
— *Profit*: How much profit are we expected to make in the following week? The profit is calculated after excluding the advertisement cost, meaning: $Revenue = Cost + Profit$
— *#Keywords Used*: This value counts the number of keywords which were selected in the optimal solution
— *Average Bid*: The average value of the bid (or *MaxCPC*) of every selected keyword of the optimal solution

The above output is the average result solution of the applied budget optimization for a future 40th week of the advertising campaign. This experiment is using a simulation and we make here the assumption that the metrics are computed as if CTR, clicks, costs, and impressions were maintained the same for each $(k, b)$ choice in the future. We first run budget optimization for different values of the available budget. In Table V, we present the average results of 30 executions for the four scenarios with budgets of 50, 100, 200, 400, and 600 units (euros).

Table V. Budget optimization evaluation results

| Budget = 50 | Clicks | Cost | Profit | #Keywords Used | AverageBid |
|---|---|---|---|---|---|
| NoPredProfit | 60 | 49.94 | 219.51 | 24 | 1.49 |
| NoPredTraffic | 61 | 49.93 | 206.22 | 23 | 1.43 |
| PredProfit | 82.36 | 49.90 | 317.1 | 16 | 1.37 |
| PredTraffic | 86.51 | 49.88 | 274.81 | 18 | 1.42 |
| **Budget = 100** | **Clicks** | **Cost** | **Profit** | **#Keywords Used** | **AverageBid** |
| NoPredProfit | 108 | 99.93 | 374.98 | 25 | 1.48 |
| NoPredTraffic | 109 | 99.92 | 356.44 | 26 | 1.44 |
| PredProfit | 130.80 | 99.87 | 467.86 | 20 | 1.41 |
| PredTraffic | 134.21 | 99.92 | 364.53 | 19 | 1.46 |
| **Budget = 200** | **Clicks** | **Cost** | **Profit** | **#Keywords Used** | **AverageBid** |
| NoPredProfit | 197 | 199.87 | 621.32 | 56 | 1.55 |
| NoPredTraffic | 200 | 199.90 | 582.21 | 54 | 1.50 |
| PredProfit | 236.94 | 199.86 | 787.63 | 31 | 1.42 |
| PredTraffic | 248.60 | 199.85 | 638.13 | 32 | 1.43 |
| **Budget = 400** | **Clicks** | **Cost** | **Profit** | **#Keywords Used** | **AverageBid** |
| NoPredProfit | 333 | 389.61 | 798.90 | 98 | 1.61 |
| NoPredTraffic | 340 | 399.92 | 791.93 | 102 | 1.63 |
| PredProfit | 425.74 | 399.82 | 1313.99 | 54 | 1.51 |
| PredTraffic | 447.42 | 399.90 | 1191.51 | 45 | 1.45 |
| **Budget = 600** | **Clicks** | **Cost** | **Profit** | **#Keywords Used** | **AverageBid** |
| NoPredProfit | 333 | 389.60 | 798.90 | 97 | 1.61 |
| NoPredTraffic | 343 | 405.16 | 795.28 | 107 | 1.63 |
| PredProfit | 607.74 | 599.84 | 1645.60 | 70 | 1.56 |
| PredTraffic | 622.69 | 599.82 | 1569.21 | 68 | 1.52 |

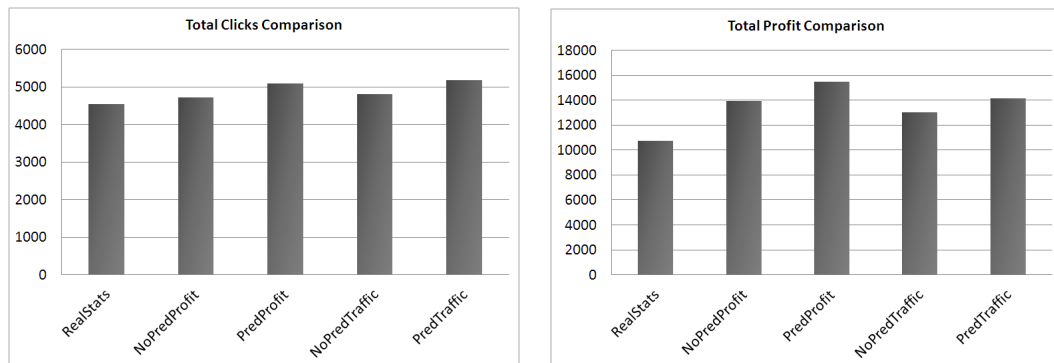

Fig. 7.   Best solutions

We notice the following on all tests: The methods that were using prediction outperform the simple GA ones. Optimization for profit always produces more profit than optimization for traffic, as expected. Optimization for traffic always produces more clicks than optimization for profit, as expected. We notice that the Average Bid increases along with the available budget. This is because the cheaper (cost-efficient) keywords are running out, so we have to use more costly ones. All solutions deplete their budget unless there are no more keywords left or the keywords left are not profitable. In the case of budget=600, when optimizing for traffic without prediction, we reach the limit of how many clicks can be made, and therefore our solution produces the maximum cost (405.16), which is less than the budget (600). This solution also uses all available keyword options (107 in size). In the case of budget=400 and budget=600, when optimizing for profit without prediction, we reach an upper limit of the profit, so the budget is not depleted. Not all keywords are used in this case because not all keywords are profitable. In the cases of small budgets, we notice that optimizing for profit generates almost as much traffic as optimizing for traffic. This could mean that keywords that generate more profit are more relevant, hence they are clicked more often.

### 6.2.2. Genetic algorithm performance on finding the best solutions for MCKP.

In the second experiment, we apply the genetic algorithm to evaluate the hypothesis of choosing the optimal keyword-bid combination of each week. The input weekly budget for our scenarios is the corresponding actual weekly cost of the campaign. For each week, the input keyword options for the genetic algorithm are the actual tested keywords and bids for the specific week. Each scenario output is the average result of five executions of the genetic algorithm. In Figure 7 we present the results for total traffic and profit comparison, where we notice that our method finds in total the most profitable keywords for both traffic and profit maximization cases.

### 6.2.3. Genetic algorithm performance on optimizing next week's performance.

In the third experiment, we test the expected weekly performance of each of our methodology scenarios towards the actual campaign weekly performance. For estimating performance of week $i$, the genetic algorithm takes into consideration the statistics from weeks 1 to $i - 1$, resulting in a "leave-one-out" cross-validation-like process. The training set is the actual statistic set from week 1 to $i - 1$ and the testing set is the actual statistic set of week $i$. For example, the input features for the optimal keywords and bids of the 20th week are the collected statistics from weeks 1 to 19. The purpose of this evaluation is to find solutions that achieve higher weekly performance than the actual one. Each scenario output from the budget optimization process is the average result of 10 executions of the genetic algorithm. The input weekly budget for our scenarios is a bit higher (1-2 euros) than the corresponding actual weekly cost of the campaign, assuming without having the actual information, that on average the budget is not completely depleted. As we present in Figure 8, in the case of traffic maximization as the advertising goal, our two methods which use prediction, surpass the real results. In this experiment, the optimization process had started after the 4th week, because the advertiser until the 3rd week had been testing very few keyword options (3-4) and the GA needs more testing data to perform a valid optimization. The important observation here compared with the stronger performance of the previous experiment was the use of much older and thus outdated data that did not correspond to valid receiving impressions and clicks in the $i$th week. The Impressions Prediction module had a major contributed role in the calculation of more up-to-date data because it achieved to capture current external factors and conditions of the ad auction. Thus, the methods that were using prediction outperformed the other ones.

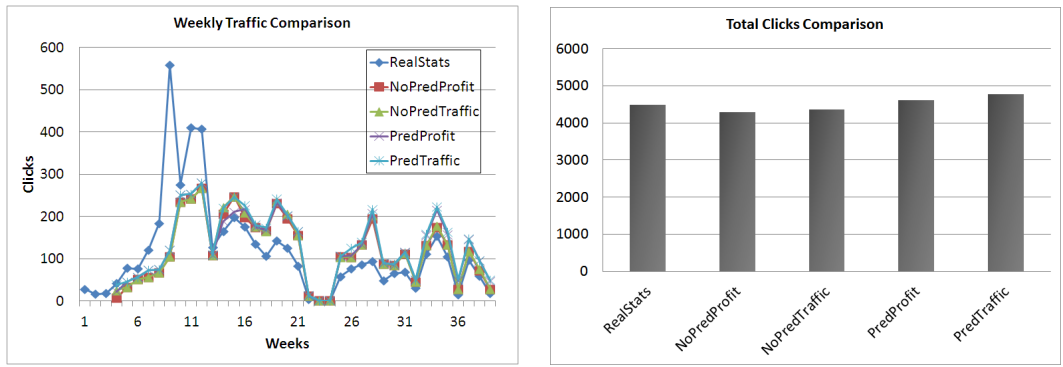

Fig. 8.   Optimization for next week

### 6.2.4. Comparison of parallel competing campaigns (Automated compared to Manual).

In the fourth experiment, we create Google AdWords campaigns for two companies; Client1 is a company that offers web developing solutions (a highly competitive field for online advertising) and Client2 is a company that offers aluminum railing and fencing products. For each company we create one manual and one automated campaign. Each automated campaign is created semi-automatically by our system (the only intervention is the parameter input of daily budget, account credentials, period of active campaign, and keywords). We set our automated campaigns for traffic maximization as the advertising goal. We use for each manual and automated campaign the same keywords and the same budget in order to test only the monitoring and optimization process. In this experiment, we do not use impressions prediction, only the real values case scenario (due to limited budget for further experiments at that time). In Figures 9 and 10, we present the final results after a period of 17 days. In the case of Client2, the automated campaign achieved higher performance in total traffic than the manual one. In the case of Client1, the automated achieved a slightly lower performance than the manual one. In both cases, the automated campaigns achieved better placement in the advertising slots than the manual ones, as well as lower prices for average cost-per-click.

## 7. CONCLUSIONS AND FUTURE WORK

In this paper we proposed a system that, given a landing page in the context of online advertising for products and services promotion, automatically extracts and suggests keywords for web advertising campaigns as well as automatically generates adver-

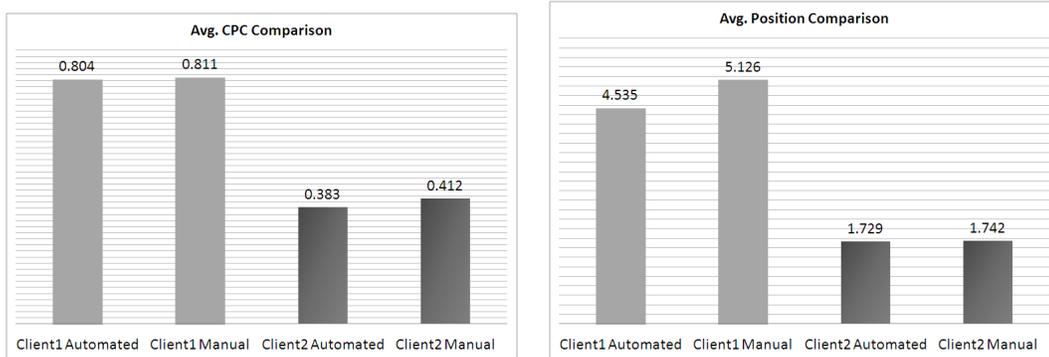

Fig. 9.   Automated compared to Manual Campaigns

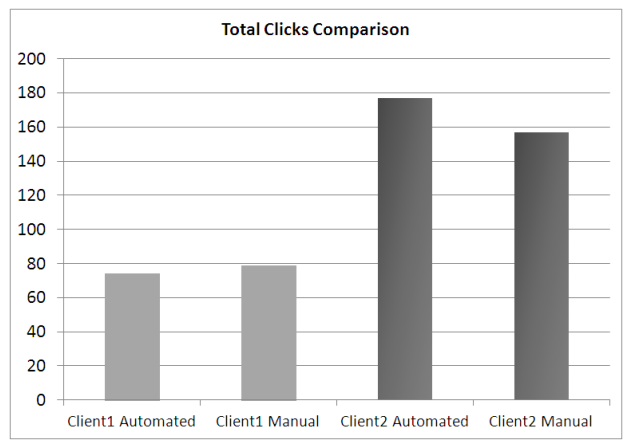

Fig. 10.   Total Clicks Comparison

tisement texts. In addition, we have proposed an architecture and a prototype frame-
work for automated advertising campaign development, monitoring, and optimization,
putting emphasis on the campaign creation, management, and budget optimization
modules. We articulated the optimization problem as a multiple-choice knapsack for
which we find the most profitable combination of keywords and their bids. We approx-
imated the solution capitalizing on a genetic algorithm for budget optimization with
multiple keyword options. We also proposed the use of keyword statistics to predict
keyword behavior using multiple linear regression. Both a. the use of a *genetic algo-
rithm* and b. *impressions prediction* for this type of problem form innovative solutions
with respect to existing literature. Proof of concept was given with the implementa-
tion of the proposed architecture. The implemented system was extensively evaluated
with data from a real website providing insight on how budget optimization works
with and without *keyword impressions prediction*. We have also tested the integrated
system for two real world web sites and compared our automated campaigns towards
the manually created ones. The budget optimization problem, even though it is an NP-
hard problem, has been practically solved by modeling it as a multiple-choice knapsack
problem.

In this way, our contributions regarding the improvement of the advertising cam-
paign development process consist in:

— Automating the task of finding the appropriate keywords
— Recommending multiword terms (n-grams) with high specificity without the need to capitalize on usage data such as query and web traffic logs
— Generating fast snippets of ad texts
— Proposing a method of overall campaign optimization
— A fully developed system with convincing experimentation on real world data from various thematic areas

Using the search result snippets for the process of keyword suggestion has helped a lot to retrieve faster the proper information rather than crawling actual documents. It was also a helpful mean to keep the trends and thus retrieving trending topics at a specific time. Also, searching result snippets from queries on twitter search and tags can be helpful due to the compact nature of twitter messages. They can help in filtering out irrelevant or general information, while mining market trends.

A further extension on our system can be the expansion of the ad creative generation component. The creation of specialized ad text will be based on previous work and research studies on paraphrasing methods, sentence extraction and compression, sentence and surface realizers, and text summarization. In combination with category specific templates which will be filled with the product characteristics, such as name, price, location, etc., the system will generate ad text for the advertisements of the campaign. The above features will be extracted from customer's web page primarily. We will experiment with 3 approaches for generating the sentences that will describe our advertisement:

(1) The advertiser can manually give as input an indicative phrase or sentence that thinks it describes best the promoted product and use paraphrase of this sentence to generate candidate ad texts. Paraphrasing methods recognize, generate, or in our case extract phrases, sentences, or longer natural language expressions that convey almost the same information [Androutsopoulos and Malakasiotis 2010]. Filtering the generated sentences with the keywords that we are bidding on, we can keep only the most relevant.

(2) We can use automatic text summarization techniques for summarizing the content of the given landing page and then generate paraphrases from the resulted sentences [Choi et al. 2010; Zhang et al. 2004]. Taking into consideration the previous described limitations of ad-texts lengths we could use sentence compression such as the method described in [Galanis and Androutsopoulos 2010].

(3) In the final ideal approach, having only as input the generated keywords from the previous process we could generate indicative phrases using them and then paraphrase them, resulting to satisfactory ad texts.

As a future challenge for more attractive advertisements, the system could take into account sentiment analysis as well.

Regarding the optimization process in future work we plan to:

— Test more methods and apply, for example, Expectation Maximization [Zacharouli et al. 2009] techniques to cluster the keyword data
— Additionally, Hidden Markov Model [Bilmes 1998] could be tested to see if there are any transitions in keyword state that could be predicted
— Take into consideration location features [Lymberopoulos et al. 2011]
— Explore the potentials of a reinforcement learning method such as Contextual Bandit Learning [Li et al. 2010] in order to exploit the various campaign features

Currently, we are already working toward alternate bidding strategies and prediction of clicks using regression trees with some preliminary but promising results. Regard-ing the overall system performance evaluation, we aim to conduct larger and combined experiments, testing concurrently both components.

## 8. ACKNOWLEDGEMENTS

The research of S. Thomaidou is co-financed by the European Union (ESF) and Greek national funds via Program Education and Lifelong Learning of the NSRF - Program: Heracleitus II. Prof. M. Vazirgiannis is partially supported by the DIGITEO Chair grant LEVETONE in France.